\documentclass{appolb}
\makeatletter
\@ifundefined{normalsize}{\input{size11.clo}}{}
\makeatother
\usepackage{amsmath,amssymb}
\usepackage{graphicx}
\usepackage{booktabs}
\usepackage{cite}
\usepackage{url}
\usepackage{microtype}
\usepackage[hidelinks]{hyperref}

\makeatletter
\def\Out@received{}
\makeatother
\headauthor{}
\headtitle{}
\makeatletter
\renewcommand\subsection{\@startsection{subsection}{2}{\z@}%
  {11pt plus 3pt minus 2pt}{5pt plus 1pt minus 1pt}%
  {\normalfont\normalsize\it\centering}}
\long\def\@makecaption#1#2{%
  \vskip\abovecaptionskip
  \sbox\@tempboxa{\small #1. #2}%
  \ifdim\wd\@tempboxa>\hsize
    {\small #1. #2\par}%
  \else
    \global\@minipagefalse
    \hbox to .993\hsize{\hfil\box\@tempboxa\hfil}%
  \fi
  \vskip\belowcaptionskip}
\def\@openbib@code{\setlength\itemsep{0pt plus 0.5pt}%
  \def\small{\@setfontsize\small\@xpt{11.7}}}
\makeatother

\newcommand{\LHmuC}{LH$\mu$C\xspace}
\newcommand{\muLHC}{$\mu$LHC\xspace}
\newcommand{\LHeC}{LH$e$C\xspace}
\newcommand{\muTRISTAN}{$\mu$TRISTAN\xspace}
\newcommand{\mup}{$\mu^+$\xspace}
\newcommand{\lumi}{cm$^{-2}$s$^{-1}$\xspace}
\newcommand{\doi}[1]{\textsc{doi}: \href{https://doi.org/#1}{\nolinkurl{#1}}}
\newcommand{\arXiv}[2]{arXiv: \href{https://arxiv.org/abs/#1}{\nolinkurl{#1}}\ifx&#2&\else\ \texttt{[#2]}\fi}
\newcommand{\urlr}[1]{\textsc{url}: \url{#1}}
\newcommand{\ws}{Talks at this workshop}

\begin{document}

\title{LH$\mu$C Concept: Energy-Frontier Antimuon--Proton\\
Collisions at CERN%
\thanks{Presented at the Workshop on Ultra-cold Muons for the LHC, Krak\'ow, Poland, 24--26 September 2026, \url{https://indico.cern.ch/event/1681458/}}%
}
\author{Arda Goroglu$^{a}$, Saleh Sultansoy$^{a,b}$
\address{$^a$TOBB University of Economics and Technology,\\
S\"o\u{g}\"ut\"oz\"u Cad.~43, 06560 Ankara, T\"urkiye\\[1pt]
$^b$Institute of Physics, Ministry of Science and Education,\\
H.~Javid Ave.~131, AZ1143 Baku, Azerbaijan}
}
\maketitle

\begin{abstract}
The concept of the LH$\mu$C, an energy-frontier antimuon--proton collider based on the HL-LHC, is presented. After recalling why energy-frontier lepton--hadron colliders are indispensable as QCD explorers and reviewing the history of muon--proton collider proposals, two ways of realising $\mu p$ collisions at the LHC are compared: with muon-collider beams and with ultra-cold $\mu^+$ beams as proposed for $\mu$TRISTAN. For the latter, a $\mu$TRISTAN-type booster and the re-use of the LH$e$C energy-recovery-linac tunnel are discussed; colliding 1~TeV antimuons with 7~TeV protons gives $\sqrt{s}=5.3$~TeV and peak luminosities of $8.8\times10^{33}$ and $7.2\times10^{33}$~cm$^{-2}$s$^{-1}$, respectively. The detector concept and the physics potential, including searches for colour-octet muons and R-parity-violating squarks, are discussed together with the main challenges.
\end{abstract}

%% =================== 1 ===================
\section{Introduction}

Lepton--hadron collisions have played an exceptional role in revealing the internal structure of matter: the proton form factors were first observed in electron scattering, quarks in deep inelastic electron scattering at SLAC, and the EMC effect in deep inelastic muon scattering at CERN~\cite{hist}. HERA, the first and still unique electron--proton collider, extended the kinematic region by orders of magnitude, shed light on the basics of quantum chromodynamics (QCD) and provided the parton distribution functions (PDFs) needed to interpret the Tevatron and LHC data~\cite{hera}.

\begin{figure}[t]
\centering
\includegraphics[width=0.84\textwidth]{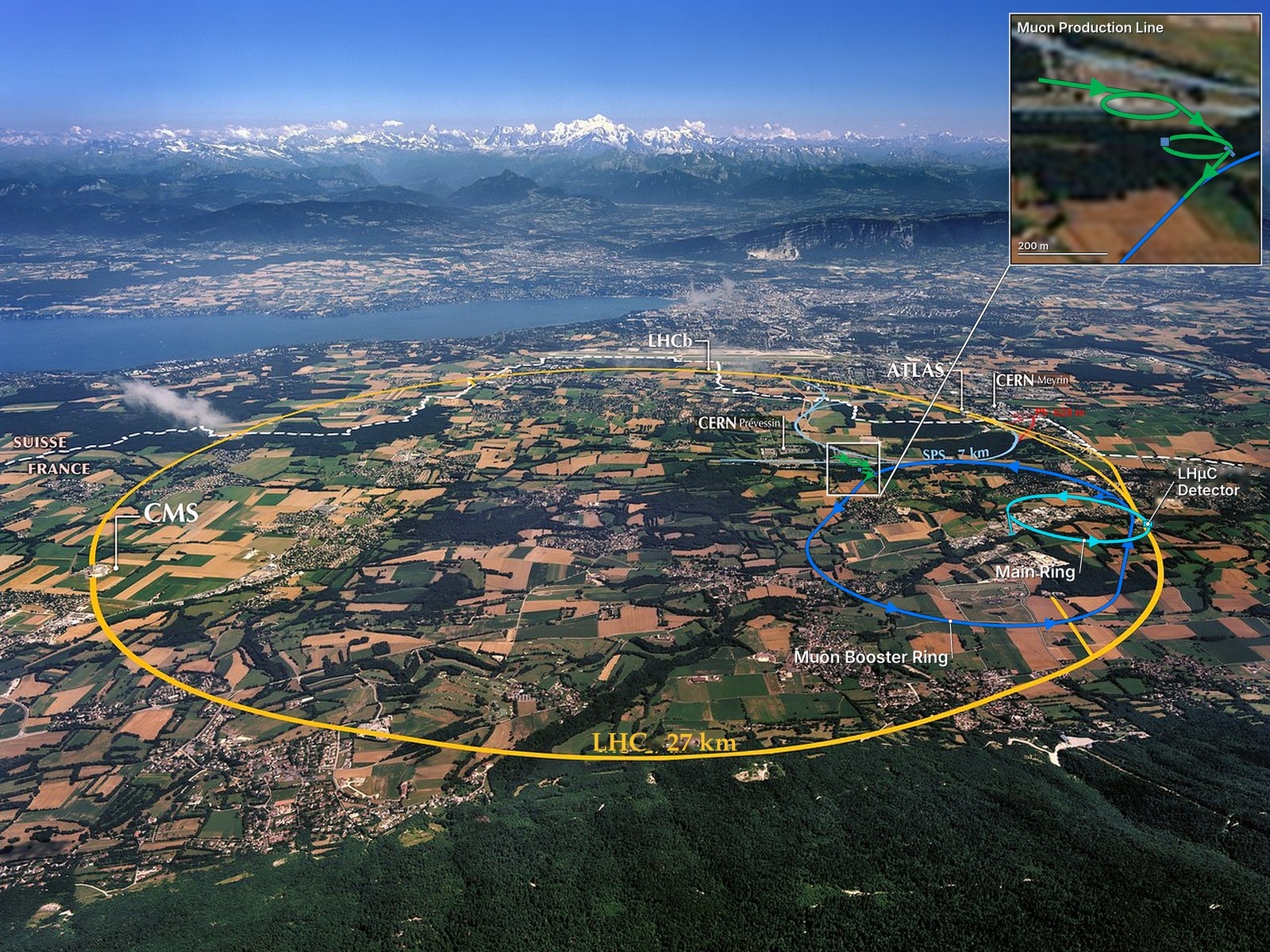}
\caption{Bird's-eye view of the \LHmuC inspired from \muTRISTAN.}
\label{fig:birdseye}
\end{figure}

While the electroweak part of the Standard Model (SM) was completed with the discovery of the Higgs boson~\cite{higgs}, large gaps remain in its QCD part: confinement has not been derived from the QCD basics, and hadronisation and nuclearisation are not yet understood. Notably, the Higgs mechanism is responsible for only about 1\% of the mass of the visible Universe: \mbox{$m_u\approx2.2$~MeV}, \mbox{$m_d\approx4.7$~MeV} and \mbox{$m_e\approx0.511$~MeV}, whereas \mbox{$m_p\approx938$~MeV}, \mbox{$m_\alpha\approx3727$~MeV}, and visible matter consists of about 74\% hydrogen and 24\% helium. The remaining 99\% originates from the strong interactions.

The region of small Bjorken $x$ ($x<10^{-4}$) at high $Q^2$ ($Q^2>10$~GeV$^2$), which is very important for understanding the QCD basics, was not covered by HERA, and the centre-of-mass energy of the EIC~\cite{eic} is several times lower than that of HERA. The \LHeC~\cite{lhec2012,lhec2021} will touch this region, whereas muon--hadron colliders can reach much higher energies. TeV-scale lepton--hadron colliders are also needed to provide PDFs for an adequate interpretation of the HL-LHC (as well as FCC-$hh$ and SppC) data. Lepton--hadron colliders are thus unique QCD explorers: Higgs physics requires the highest luminosity, whereas the QCD basics require the highest energy. Table~\ref{tab:lp} compares the lepton--proton colliders that could operate in the 2030s with HERA: the \muLHC~\cite{ptep}, sketched in Fig.~\ref{fig:birdseye}, would provide a 4.5 times higher centre-of-mass energy than the \LHeC, at a comparable luminosity.

\begin{table}[htb]
\caption{Lepton--proton colliders of the 2030s compared with HERA; ERL\,20 is the single-pass option of Ref.~\cite{andre}.}
\label{tab:lp}
\centering
\begin{tabular}{lccc}
\toprule
Collider & $\sqrt{s}$ [TeV] & $\mathcal{L}$ [cm$^{-2}$s$^{-1}$] & $L_{\rm int}$ [fb$^{-1}$]\\
\midrule
HERA                       & 0.319 & $7.5\times10^{31}$ & 0.8 (total)\\
EIC                        & 0.105 & $1.0\times10^{34}$ & 100/year\\
\LHeC (ERL\,20)            & 0.748 & $6.0\times10^{33}$ & 60/year\\
\LHeC (ERL\,50)            & 1.18  & $5.0\times10^{33}$ & 50/year\\
\muLHC (\muTRISTAN booster)& 5.29  & $8.8\times10^{33}$ & 88/year\\
\muLHC (ERL booster)       & 5.29  & $7.2\times10^{33}$ & 72/year\\
\bottomrule
\end{tabular}
\end{table}

%% =================== 2 ===================
\section{Muon--proton colliders: from early proposals to the \boldmath$\mu$LHC}

Muon--proton colliders were proposed three decades ago as an alternative to linac--ring type electron--proton colliders~\cite{shiltsev,ginzburg,cheung98,lrreviews}. For a muon ring tangential to the Tevatron, luminosities of $10^{32}$--$10^{33}$~\lumi were estimated~\cite{shiltsev}, and the potential for leptoquark and contact-interaction searches was explored~\cite{cheung98}. As the synchrotron radiation of muons is $(m_\mu/m_e)^4\approx2\times10^9$ times weaker than that of electrons, energy-frontier muon--hadron colliders can be ring--ring machines, unlike the linac--ring type electron--hadron ones. Later, FCC- and SppC-based $\mu p$ and $\mu A$ colliders~\cite{acar17}, LHC- and FCC-based $\mu p$ colliders~\cite{cheungwang} and a muon--ion collider at BNL~\cite{acosta} were proposed; see Ref.~\cite{ketenoglu22} for a review.

\emph{Muon-collider-based option.} A muon collider~\cite{mc} built tangentially to the LHC would provide $\mu p$ collisions with the HL-LHC (HE-LHC) protons~\cite{kaya22} at $\sqrt{s}=4.58$--9.16 (12.7)~TeV and luminosities of $(0.6$--$2.1)\times10^{33}$~\lumi, extending, \emph{e.g.}, the discovery reach for colour-octet muons from 3.5~TeV at the HL-LHC to about 6.5 (9)~TeV~\cite{kaya22}. This option, however, has to await the solution of the muon cooling problem, with a cooling demonstrator foreseen in the 2030s~\cite{mc}.

\emph{Ultra-cold antimuon option.} While a low-emittance $\mu^-$ beam has not been achieved yet, there is an established technology to create a low-emittance $\mu^+$ beam using ultra-cold muons, developed for the muon $g-2$/EDM experiment at J-PARC~\cite{abe}. The \muTRISTAN proposal~\cite{hamada} showed how such a source can be intensified by orders of magnitude and the beam accelerated to 1~TeV without ionisation cooling. Antimuon rings tangential to the FCC-$ee$/CEPC and to the LHC, Tevatron, FCC and SppC were then considered~\cite{akturk}, leading to the HL-LHC-based \muLHC~\cite{ptep}, which is now developed further by the HELP4CERN Study Group as the \LHmuC~\cite{help4cern,kpdis,talks}.

%% =================== 3 ===================
\section{LH\boldmath$\mu$C: antimuon complex and luminosity}

\subsection{From ultra-cold muons to a 1~TeV antimuon beam}

\begin{figure}[htb]
\centering
\includegraphics[width=0.66\textwidth]{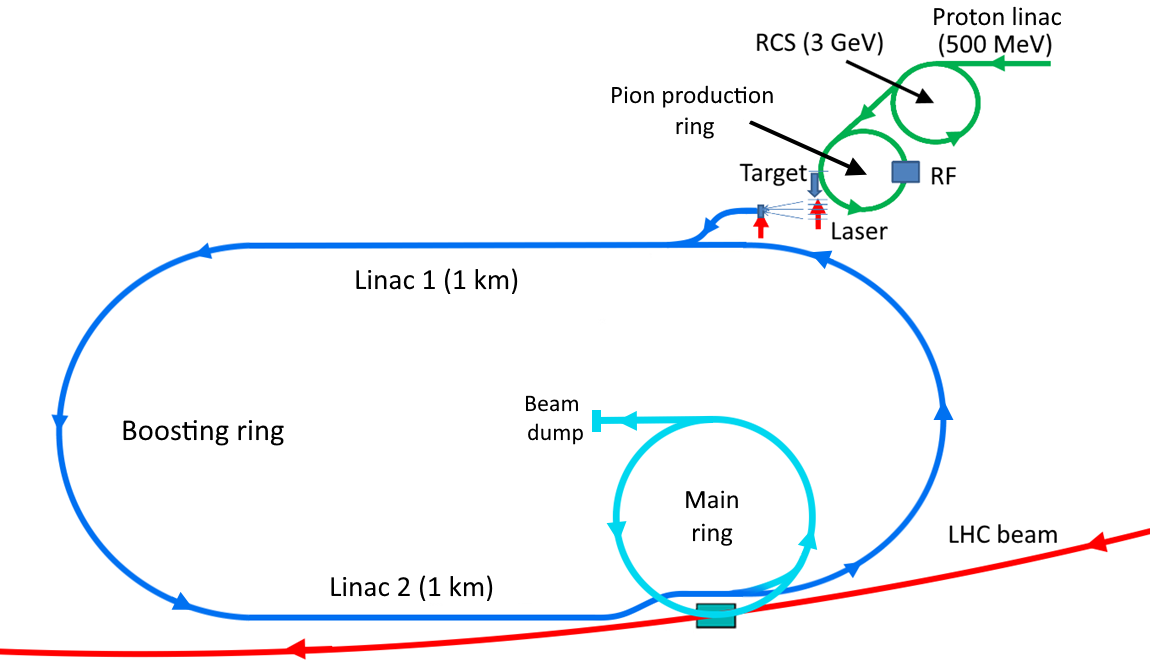}
\caption{Schematic view of the \LHmuC (\muLHC) with the ERL-based booster: the \mup booster (``boosting ring'') and the main ring are integrated within the \LHeC ERL tunnel, similarly to the \muTRISTAN layout (from Ref.~\cite{ptep}).}
\label{fig:layout}
\end{figure}

The \LHmuC antimuon complex (Fig.~\ref{fig:layout}) consists of six stages~\cite{hamada,ptep,daglws}: (i)~a 3~GeV, 2~MW rapid-cycling synchrotron operating at 50~Hz; (ii)~a pion-production ring, in which each proton bunch crosses a 10~mm graphite target 40 times, yielding 0.016 pions per proton; (iii)~an ultra-cold \mup source, where the pions are stopped in a multi-layer tungsten target and the thermal muonium is laser-ionised; (iv)~a linac re-accelerating the \mup to 212~MeV with a normalised emittance $\varepsilon_n=4~\mu$m, in trains of 40 bunches of $4.5\times10^{10}$ \mup; (v)~a booster accelerating them to 1~TeV in about 1~ms; and (vi)~a 3~km main ring tangential to the LHC, where the \mup collide with the LHC protons at one interaction point (IP). The ultra-cold \mup yield is thus about $10^5$ times higher than at the present J-PARC MLF beamline, and the low emittance is achieved without ionisation cooling, so that a $\mu^+$--hadron collider can be realised well before a $\mu^+\mu^-$ collider. The source and the low-energy acceleration are being actively developed~\cite{aritome,wssource,wstarget}.

\subsection{Booster options}

\begin{figure}[htb]
\centering
\includegraphics[width=0.76\textwidth]{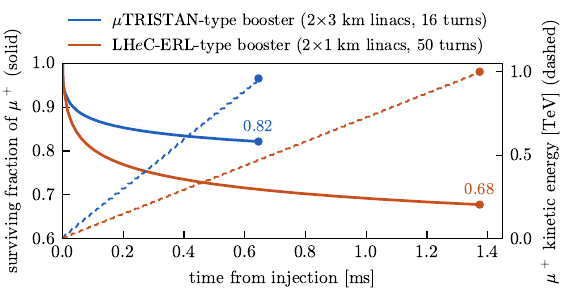}
\caption{Surviving fraction (solid lines) and kinetic energy (dashed lines) of the \mup \emph{vs.} time from injection at 212~MeV for the two booster options (this work, parameters of Ref.~\cite{ptep}).}
\label{fig:survival}
\end{figure}

Two booster options were studied in Ref.~\cite{ptep}. The \muTRISTAN-type booster~\cite{hamada} is a racetrack with two 3~km linacs of 30~GeV each and 1~km radius arcs; the \mup reach 960~GeV after 16 turns, in about 0.65~ms. In the second option, proposed by F.~Zimmermann, the \LHeC energy-recovery-linac (ERL) tunnel~\cite{lhec2021} --- a similar racetrack with two 1~km linacs providing 10~GeV per pass --- is re-used (Fig.~\ref{fig:layout}), and 1~TeV is reached after 50 turns, in about 1.4~ms. This allows for a natural staging: the ERL is first used for $ep$ and $eA$ collisions, possibly starting with the single-pass 20~GeV \LHeC~\cite{andre}, and the \mup booster is installed later in the same tunnel. Figure~\ref{fig:survival} shows the \mup surviving fraction and energy as functions of time for both options: our calculation gives a survival of 0.82 and 0.68, in agreement with the values of 0.82 and 0.67 of Ref.~\cite{ptep},\footnote{These values (second arXiv version of Ref.~\cite{ptep}) supersede the original ones (0.38 and 0.16), which were affected by an incorrect muon mass; the luminosities of Sect.~\ref{sec:lumi} are thus about two and four times larger.} \emph{i.e.} $3.7\times10^{10}$ and $3.0\times10^{10}$ \mup per bunch at the top energy.

The rapid acceleration of a multi-bunch train within the muon lifetime is the most critical challenge of the booster. Unlike the hybrid rapid-cycling synchrotron of the Muon Collider~\cite{mc}, a vertical-excursion fixed-field accelerator (vFFA)~\cite{topp} needs no magnet ramping and keeps the 40-bunch train intact, and cascaded vFFA arcs could be used in the \LHeC tunnel~\cite{ptep}.

\subsection{Main ring and luminosity}\label{sec:lumi}

The 40 \mup bunches circulate in the 3~km main ring with a spacing of 250~ns --- every tenth LHC bucket --- giving a collision frequency $f_c=4$~MHz. At 1~TeV the muon lifetime is 20.8~ms ($\approx2000$ turns); with refilling at 50~Hz, the average \mup population is $1-e^{-1}\approx0.63$ of the injected one. For the ``ERL60-upgraded'' \LHeC proton beam~\cite{lhec2012} and matched beam sizes (Table~\ref{tab:param}), the luminosity $\mathcal{L}=f_cN_\mu N_p/(4\pi\sigma_x\sigma_y)$, calculated with the AloHEP software~\cite{alohep} including the hourglass and pinch effects and the muon decay, is $8.8\times10^{33}$ and $7.2\times10^{33}$~\lumi for the two options. As only every tenth LHC bunch collides, fewer but more intense proton bunches~\cite{zimmws} would substantially increase the luminosity: about $2\times10^{34}$~\lumi was estimated with a 4~MW proton driver, even for only 5\% of the \muTRISTAN \mup yield~\cite{kpdis}.

\subsection{Interaction region and detector}

\begin{figure}[t]
\centering
\includegraphics[width=0.54\textwidth]{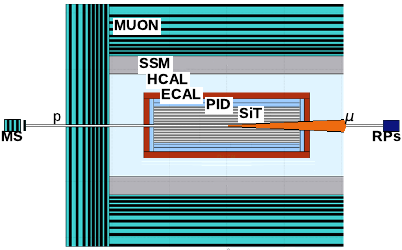}
\caption{Detector concept for $\mu^+p$ collisions~\cite{ptep}: silicon tracker (SiT), particle identification (PID), calorimeters (ECAL, HCAL), solenoid (SSM), muon system (MUON), Roman pots (RPs) and far-backward muon spectrometer (MS); the tungsten nozzle is shown in orange.}
\label{fig:detector}
\end{figure}

Owing to the large \mup bunch spacing, a very small crossing angle suffices to avoid parasitic collisions~\cite{kpdis}. The detector concept~\cite{ptep} (Fig.~\ref{fig:detector}), similar to that for the muon--ion collider~\cite{acosta}, has an asymmetric layout complemented by Roman pots and a muon spectrometer far from the IP. The background from \mup decays is suppressed by a tungsten nozzle on the incoming-\mup side, which limits the acceptance; 1~TeV muons traversing it lose on average about 19\% of their energy~\cite{ptep}. These issues, as well as TeV-muon and near-beam detectors, were discussed at this workshop~\cite{wsdetector}.

\begin{table}[t]
\caption{Main parameters of the \LHmuC (\muLHC) for the two booster options~\cite{ptep}; the \mup bunch population is given at injection into the main ring.}
\label{tab:param}
\centering
\begin{tabular}{lcccc}
\toprule
Booster option & \multicolumn{2}{c}{\muTRISTAN-type} & \multicolumn{2}{c}{ERL-based}\\
\midrule
$\sqrt{s}$ [TeV] & \multicolumn{2}{c}{5.3} & \multicolumn{2}{c}{5.3}\\
$\mathcal{L}$ [$10^{33}$~cm$^{-2}$s$^{-1}$] & \multicolumn{2}{c}{8.8} & \multicolumn{2}{c}{7.2}\\
\midrule
Beam parameters & $\mu^+$ & $p$ & $\mu^+$ & $p$\\
\midrule
Energy [TeV] & 1 & 7 & 1 & 7\\
Bunch population [$10^{10}$] & 3.7 & 22 & 3.0 & 22\\
Norm. emittance [$\mu$m] & 4 & 2 & 4 & 2\\
RMS beam size at IP, $\sigma_{x,y}$ [$\mu$m] & 4.3 & 4.3 & 4.3 & 4.3\\
Tune shift, $\xi_{x,y}$ [$10^{-4}$] & 600 & 22 & 600 & 18\\
\bottomrule
\end{tabular}
\end{table}

%% =================== 4 ===================
\section{Physics potential}

Multi-TeV muon--hadron colliders have a huge potential for SM (especially QCD) and beyond-the-SM (BSM) physics, notably for muon-related phenomena such as excited muons, colour-octet muons, leptoquarks and contact interactions~\cite{cheung98,cheungwang,acosta,ketenoglu22,bsmmup}; nuclear options, two-photon, Higgs-pair and neutrino physics were discussed at this workshop~\cite{daglws,wsphysics}.

\begin{figure}[t]
\centering
\begin{minipage}[b]{0.44\textwidth}\centering
\includegraphics[width=\textwidth]{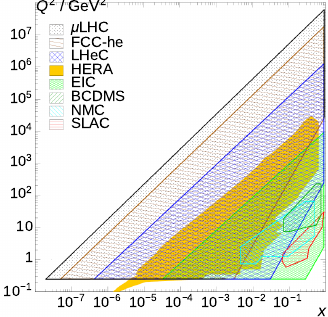}\\[2pt]
{\small (a)}
\end{minipage}\hfill
\begin{minipage}[b]{0.49\textwidth}\centering
\includegraphics[width=\textwidth]{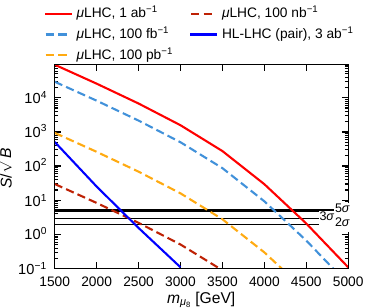}\\[2pt]
{\small (b)}
\end{minipage}
\caption{(a) Coverage of the kinematic plane in deep inelastic lepton--proton scattering by the \muLHC, compared with the FCC-$eh$, \LHeC, HERA, EIC and fixed-target experiments. (b) Statistical significance of the colour-octet muon signal \emph{vs.} its mass for single production at the \muLHC with various integrated luminosities, and for pair production at the HL-LHC with 3~ab$^{-1}$. Both adapted from Ref.~\cite{ptep}.}
\label{fig:phys}
\end{figure}

\emph{QCD at small $x$.} Since $Q^2=xys$ with $y\le1$, at $\sqrt{s}=5.3$~TeV values as low as $x=3.6\times10^{-7}$ are accessible at $Q^2=10$~GeV$^2$, compared with $1.0\times10^{-4}$ at HERA and $7.2\times10^{-6}$ at the \LHeC (ERL\,50); at $x=10^{-4}$, $Q^2$ extends up to 2800~GeV$^2$, compared with 10 and 140~GeV$^2$, respectively~\cite{ptep}. As seen in Fig.~\ref{fig:phys}(a), the \muLHC covers a much wider range of small $x$ and high $Q^2$ than the \LHeC and even the FCC-$eh$, giving unique access to non-linear parton dynamics~\cite{kpdis,wsphysics}.

\emph{Higgs boson.} Higgs bosons are produced via $WW$ ($ZZ$) fusion with a cross section of about 0.39 (0.075)~pb, \emph{i.e.} $3.4\times10^4$ ($6.6\times10^3$) Higgs bosons per year for the \muTRISTAN-type option~\cite{ptep}, an order of magnitude more than at the \LHeC, which will help to determine the Higgs couplings to the $W$ and $Z$ bosons~\cite{wsphysics}.

\emph{Colour-octet muons.} In composite models, each SM lepton is accompanied by a colour-octet partner, the leptogluon~\cite{goncalves}; the colour-octet muon $\mu_8$ shares the electroweak quantum numbers of the muon. Its interaction with leptons $\ell$ and gluons reads
\begin{equation}
\mathcal{L}=\frac{1}{2\Lambda}\left\{\bar{\ell}_8^{\,\alpha}\,g_s G^{\alpha}_{\mu\nu}\sigma^{\mu\nu}\left(\eta_L\ell_L+\eta_R\ell_R\right)+{\rm h.c.}\right\},
\label{eq:mu8}
\end{equation}
where $\Lambda$ is the compositeness scale (set to $m_{\mu_8}$) and $\eta_{L,R}$ are chirality factors. At the LHC colour-octet muons are pair-produced, whereas at $\mu p$ colliders they are produced resonantly, $\mu g\to\mu_8$, with a much larger cross section~\cite{kaya22,bsmmup}. A hadron-level analysis~\cite{ptep}, requiring one muon and one jet with $p_T>400$~GeV at the \muLHC and two muons and two jets with $p_T>300$ (250)~GeV for the leading (sub-leading) objects at the HL-LHC, shows that the $5\sigma$ discovery reach of about 2.3~TeV at the HL-LHC with 3~ab$^{-1}$ is extended to about 4.1 (4.4)~TeV at the \muLHC with 100~fb$^{-1}$ (1~ab$^{-1}$), see Fig.~\ref{fig:phys}(b); the ultimate HL-LHC sensitivity is matched with an integrated luminosity smaller by at least four orders of magnitude. An update gives a $5\sigma$ reach of about 3 (4.2)~TeV for a 0.5 (1)~TeV \mup beam with 250 (500)~fb$^{-1}$~\cite{talks}.

\emph{R-parity-violating supersymmetry.} The $\lambda'_{ijk}L_iQ_j\bar{D}_k$ terms of the R-parity-violating superpotential~\cite{rpv} allow squarks to be produced as $s$-channel resonances in muon--quark collisions~\cite{rpv,cheungwang}. At the \LHmuC, the neutral-current process $\mu^+d\to\tilde{u}_L\to\mu^+d$ proceeds on valence quarks, whereas the charged-current one, $\mu^+\bar{u}\to\tilde{d}_R^{\,*}\to\bar{\nu}_\mu\bar{d}$, relies on sea antiquarks and is less sensitive. For 1~TeV antimuons and 500~fb$^{-1}$, requiring exactly one muon, at least one jet and a high scalar sum of the jet transverse momenta, $\lambda'_{2j1}$ can be probed down to 0.01--0.02 for squark masses of 1--3~TeV, almost an order of magnitude beyond the current and projected LHC bounds~\cite{talks}.

%% =================== 5 ===================
\section{Challenges and conclusions}

The main challenges of the \LHmuC are: the antimuon yield (stages i--iii), whose $10^5$-fold enhancement has to be demonstrated~\cite{wssource,wstarget}; the booster magnets (stage v), where fixed-field arcs are needed~\cite{topp} and the use of high-gradient X-band linacs instead of superconducting ones deserves study; the radiation induced by neutrinos from \mup decays (stages v--vi), which, on the other hand, provide intense and collimated neutrino beams~\cite{wsphysics}; and the interaction region and detector design~\cite{wsdetector}.

In conclusion, energy-frontier lepton--hadron colliders are a must for two reasons: the clarification of the strong interactions, including the QCD basics, which are responsible for about 99\% of the mass of the visible Universe, and the provision of PDFs for an adequate interpretation of the HL-LHC (as well as FCC-$hh$ and SppC) data. Since the LHC provides the highest-energy hadron beams, CERN is the most suitable place for building such colliders. Based on the ultra-cold \mup technology, the \muLHC/\LHmuC appears to be the most realistic way to reach multi-TeV centre-of-mass energies in lepton--hadron collisions in the 2030s, with $\sqrt{s}=5.3$~TeV and luminosities approaching $10^{34}$~\lumi. In addition, it will provide a unique opportunity to investigate muon-related BSM physics. Systematic studies of its accelerator, detector and physics aspects, as pursued by the HELP4CERN Study Group~\cite{help4cern}, are therefore strongly motivated.

\section*{Acknowledgements}

We thank K.~Piotrzkowski and the Krak\'ow group for organising this workshop, and K.~Piotrzkowski for pointing out the inconsistency in the antimuon survival rates of Ref.~\cite{ptep}. We are grateful to A.~C.~Canbay, B.~Dagli, L.~Forthomme, U.~Kaya, B.~Ketenoglu, A.~Kilic, F.~Kocak, I.~Tapan, F.~Zimmermann and the HELP4CERN Study Group for fruitful discussions. This work is partially supported by T\"UB\.ITAK under grant No.~125F120.

%% =================== references ===================


\begin{thebibliography}{99}

\bibitem{hist}
R.~Hofstadter and R.~W.~McAllister. In: \textit{Phys. Rev.} 98 (1955), p.~217; E.~D.~Bloom et al. In: \textit{Phys. Rev. Lett.} 23 (1969), p.~930; M.~Breidenbach et al. In: \textit{Phys. Rev. Lett.} 23 (1969), p.~935; J.~J.~Aubert et al. (EMC). In: \textit{Phys. Lett. B} 123 (1983), p.~275.

\bibitem{hera}
M.~Klein and R.~Yoshida. ``Collider physics at HERA''. In: \textit{Prog. Part. Nucl. Phys.} 61 (2008), p.~343. \arXiv{0805.3334}{}.

\bibitem{higgs}
ATLAS Collaboration. In: \textit{Phys. Lett. B} 716 (2012), p.~1; CMS Collaboration. In: \textit{Phys. Lett. B} 716 (2012), p.~30.

\bibitem{eic}
R.~Abdul Khalek et al. ``EIC Yellow Report''. In: \textit{Nucl. Phys. A} 1026 (2022), p.~122447. \arXiv{2103.05419}{}.

\bibitem{lhec2012}
J.~L.~Abelleira Fernandez et al. (LHeC Study Group). ``A Large Hadron Electron Collider at CERN''. In: \textit{J. Phys. G} 39 (2012), p.~075001. \arXiv{1206.2913}{}.

\bibitem{lhec2021}
P.~Agostini et al. ``The Large Hadron-Electron Collider at the HL-LHC''. In: \textit{J. Phys. G} 48 (2021), p.~110501. \arXiv{2007.14491}{}; F.~Ahmadova et al. 2025. \arXiv{2503.17727}{}.

\bibitem{ptep}
D.~Akturk et al. ``$\mu$LHC: Antimuon Ring and High Luminosity-LHC-Based $\mu^+p$ Collider''. In: \textit{PTEP} 2026 (2026), 033C04. \doi{10.1093/ptep/ptag031}. \arXiv{2506.15445v2}{}.

\bibitem{andre}
K.~Andr\'e et al. ``An electron--hadron collider at the high-luminosity LHC''. In: \textit{New J. Phys.} 28 (2026), 013001. \arXiv{2503.20475}{}.

\bibitem{shiltsev}
V.~D.~Shiltsev. FERMILAB-TM-1969. 1996; V.~D.~Shiltsev. ``An asymmetric muon--proton collider: luminosity consideration''. In: \textit{Proc. PAC97}, Vancouver. 1997, p.~420.

\bibitem{ginzburg}
I.~F.~Ginzburg. ``Physics at future $ep$, $\gamma p$ (linac--ring) and $\mu p$ colliders''. In: \textit{Turk. J. Phys.} 22 (1998), p.~607.

\bibitem{cheung98}
K.~Cheung. ``Muon--proton colliders: leptoquarks and contact interactions''. In: \textit{AIP Conf. Proc.} 441 (1998). \doi{10.1063/1.56424}. \arXiv{hep-ph/9802219}{}.

\bibitem{lrreviews}
S.~Sultansoy. DESY-99-159. 1999. \arXiv{hep-ph/9911417}{}; S.~Sultansoy. In: \textit{Eur. Phys. J. C} 33 (2004), S1064; A.~N.~Akay, H.~Karadeniz and S.~Sultansoy. In: \textit{Int. J. Mod. Phys. A} 25 (2010), p.~4589.

\bibitem{acar17}
Y.~C.~Acar et al. In: \textit{Nucl. Instrum. Meth. A} 871 (2017), p.~47. \arXiv{1608.02190}{}; A.~C.~Canbay et al. In: \textit{Adv. High Energy Phys.} 2017 (2017), 4021493. \arXiv{1704.03534}{}.

\bibitem{cheungwang}
K.~Cheung and Z.~S.~Wang. ``Physics potential of a muon--proton collider''. In: \textit{Phys. Rev. D} 103 (2021), p.~116009. \arXiv{2101.10476}{}.

\bibitem{acosta}
D.~Acosta and W.~Li. In: \textit{Nucl. Instrum. Meth. A} 1027 (2022), p.~166334. \arXiv{2107.02073}{}; D.~Acosta et al. ``The potential of a TeV-scale muon-ion collider''. In: \textit{JINST} 18 (2023), P09025. \arXiv{2203.06258}{}.

\bibitem{ketenoglu22}
B.~Ketenoglu et al. ``Review of muon--proton and muon--nucleus collider proposals''. In: \textit{Mod. Phys. Lett. A} 37 (2022), 2230013. \arXiv{2206.00037}{}.

\bibitem{mc}
J.~P.~Delahaye et al. ``Muon Colliders''. 2019. \arXiv{1901.06150}{}; C.~Accettura et al. ``Towards a muon collider''. In: \textit{Eur. Phys. J. C} 83 (2023), p.~864. \arXiv{2303.08533}{}; C.~Accettura et al. ``The Muon Collider''. 2025. \arXiv{2504.21417}{}.

\bibitem{kaya22}
U.~Kaya et al. ``Luminosity and physics considerations on HL-LHC- and HE-LHC-based $\mu p$ colliders''. In: \textit{EPL} 138 (2022), p.~24002. \arXiv{1905.05564}{}.

\bibitem{abe}
M.~Abe et al. In: \textit{PTEP} 2019 (2019), 053C02. \arXiv{1901.03047}{}.

\bibitem{hamada}
Y.~Hamada et al. ``$\mu$TRISTAN''. In: \textit{PTEP} 2022 (2022), 053B02. \doi{10.1093/ptep/ptac059}. \arXiv{2201.06664}{}.

\bibitem{akturk}
D.~Akturk, B.~Dagli and S.~Sultansoy. In: \textit{EPL} 147 (2024), p.~24001. \arXiv{2403.17034}{}; D.~Akturk et al. In: \textit{Eur. Phys. J. Plus} 140 (2025), p.~317. \arXiv{2406.02647}{}.

\bibitem{help4cern}
HELP4CERN Study Group. \textit{White Paper on High Energy, High Luminosity Lepton--Hadron Colliders at CERN}. In preparation. 2026.

\bibitem{kpdis}
K.~Piotrzkowski. ``LH$\mu$C: A high energy, high luminosity muon--hadron collider at CERN''. In: \textit{Proc. DIS2026}. 2026. \arXiv{2610.35154}{}.

\bibitem{talks}
K.~Piotrzkowski. Talk at DIS2026, Bologna, 4--8 May 2026; L.~Forthomme. Talk at ICHEP2026, Natal, 30 Jul.--5 Aug. 2026.

\bibitem{daglws}
B.~Dagli et al. ``LH$\mu$C: nucleus options''. Talk at this workshop.

\bibitem{aritome}
S.~Aritome et al. ``Acceleration of Positive Muons by a Radio-Frequency Cavity''. In: \textit{Phys. Rev. Lett.} 134 (2025), p.~245001. \arXiv{2410.11367}{}.

\bibitem{wssource}
S.~Kamioka et al.; M.~Otani; Y.~Nakazawa; Y.~Takeuchi et al.; M.~Yoshida. \ws.

\bibitem{wstarget}
C.~Zhang et al.; C.~Rogers. \ws; Y.~Sakaki et al. 2026. \arXiv{2605.25580}{}.

\bibitem{topp}
M.~E.~Topp-Mugglestone. PhD thesis. University of Oxford, 2024; M.~Topp-Mugglestone et al. \ws.

\bibitem{alohep}
B.~Dagli et al. ``ALOHEP software''. In: \textit{Proc. IPAC'21}. 2021.

\bibitem{zimmws}
F.~Zimmermann. ``Potential possibilities of modifying SPS/LHC''. Talk at this workshop.

\bibitem{wsdetector}
L.~Forthomme; Y.~Yamazaki; D.~Acosta et al.; A.~Kilic et al. \ws.

\bibitem{bsmmup}
Y.~C.~Acar, U.~Kaya and B.~B.~Oner. In: \textit{Chin. Phys. C} 42 (2018), p.~083108. \arXiv{1703.04030}{}; A.~Caliskan, S.~Kara and A.~Ozansoy. In: \textit{Adv. High Energy Phys.} 2017 (2017), 1540243. \arXiv{1701.03426}{}.

\bibitem{wsphysics}
P.~Kotko et al.; M.~Kumar et al.; H.~Khanpour; X.~Lu et al. \ws; S.~Tentori. These proceedings, in preparation.

\bibitem{goncalves}
D.~Gon\c{c}alves-Netto et al. ``Looking for leptogluons''. In: \textit{Phys. Rev. D} 87 (2013), p.~094023. \arXiv{1303.0845}{}.

\bibitem{rpv}
R.~Barbier et al. In: \textit{Phys. Rept.} 420 (2005), p.~1. \arXiv{hep-ph/0406039}{}; M.~Carena et al. In: \textit{Phys. Rev. D} 62 (2000), p.~095010. \arXiv{hep-ph/0006144}{}.

\end{thebibliography}
\end{document}